\documentclass[twocolumn,twoside,slac_two]{revtex4}
\usepackage{graphicx}
\usepackage{fancyhdr}
\begin{document}

\title{Kanata optical and X-ray monitoring of Gamma-ray emitting Narrow-Line Seyfert 1 and Radio galaxies}


\author{K. Kawaguchi, Y. Fukazawa,  R. Itoh, Y. Kanda, K. Shiki, K. Takaki}
\affiliation{Department of Physical Science, Hiroshima University,
Higashi-Hiroshima, Hiroshima 739-8526, Japan}
\author{Y.T. Tanaka, M. Uemura, H.Akitaya}
\affiliation{Hiroshima Astrophysical Science Center, Hiroshima
University, Higashi-Hiroshima, Hiroshima 739-8526, Japan}

% In case co-authors are many, please use the following expression.

%\author{F. Author$^1$, S. Author$^2$}

%\affiliation{$^1$University/Institute, City, State, Postal Code Country}
%\affiliation{$^2$Colalborative University/Institute, City, State, Postal Code Country}

\begin{abstract}
 Broadband spectrum of AGN consists of
 multiple components such as jet emission and accretion disk emission. 
 Temporal
 correlation study is useful to understand emission components and
 their physical origins. We have performed optical monitoring using
 Kanata telescope for 4 radio galaxies and 6 radio-loud Narrow-Line
 Seyfert 1 (RL-NLSy1): 2 gamma-ray-loud
 RL-NLSy1s, 1H 0323+342 and PMN J0948+0022, and 4 gamma-ray-quiet
 RL-NLSy1s. 
From these results, it is suggested that RL-NLSy1s show a disk-dominant
 phase and a jet-dominant phase in the optical band, but it is not well
 correlated with brightness.
\end{abstract}

\maketitle

\thispagestyle{fancy}


\section{Introduction}

Active Galactic Nucleus (AGNs) emit electromagnetic radiation from
 radio up to TeV gamma-ray ranges. 
Spectral Energy Distribution (SED) of blazars is often dominanted 
by 2 component, synchrotron emission and Inverse Compton from a relativistic
jet. 
However, SED of misaligned radio-loud AGNs is complicated due to disk/corona
emission. 
In addition to the above two components, we
can see disk emission from near-infrared to ultraviolet bands and corona
emission in X-ray band. Because it is difficult to separate these components,
optical emission mechanism is still unclear. \\
Radio galaxy is radio-loud AGN which has a relative large viewing
angle. Thanks to high sensitive observation by Fermi Gamma-Ray
Space Telescope/ Large Area Telescope (LAT),
correlation study between optical and MeV/GeV gamma-ray bands
has become available, but correlation
between optical and X-ray bands is still unclear. \\
Narrow-Line Seyfert 1(NLSy1) is a subclass of Seyfert 1 galaxies. 
Most of NLSy1
is radio-quiet, but a few objects(~7\%) are radio-loud. Recently, Fermi-LAT detected MeV/GeV gamma-ray
emission from radio-loud NLSy1 (RL-NLSy1) and now RL-NLSy1 is a new
class of gamma-ray emitting AGNs. Radio-loud NLSy1 shows fast and strong 
variability like blazars. The most gamma-ray
bright NLSy1 PMN J0948+0022 showed minute-scale optical variability, correlated
with polarization degree\cite{itoh_2013}. This indicates that 
synchrotron emision from the jet is dominant in the optical band, 
but other study shows disk emission is
also dominant in the optical band\cite{abdo_2009}. Hence emission mechanism
in the optical band in RL-NLSy1 is still unclear.\\

\section{Observation}
We have performed optical monitor with
the Kanata optical telescope. 
We use MAXI, Swift-BAT and Fermi-LAT public data for X-ray and gamma-ray monitor.\\
We selected famous and X-ray bright objects for radio galaxies. For
RL-NLSy1, we selected gamma-ray loud objects and a few gamma-ray quiet
objects. These gamma-ray quiet objects are reported to have a
blazar-like radio structure and high brightness temperature by Komossa et
al. (2006)\cite{komossa_2006} Doi et al. (2011)\cite{doi_2011} and Doi et
al. (2012)\cite{doi_2012}. So if these gamma-ray quiet NLSy1 has a
relativistic jet, flares in the optical band are expected.

\begin{table}[t]
\begin{center}
\caption{Target lists}
\begin{tabular}{|c|c|}
\hline \multicolumn{2}{|c|}{\textbf{Radio galaxies}} \\
\hline 3C 111 & 3C 120 \\
\hline 3C 390.3 & NGC 1275 \\
\hline \multicolumn{2}{|c|}{\textbf{Gamma-ray loud NLSy1s}} \\
\hline PMN J0948+0022 & 1H 0323+342 \\
\hline \multicolumn{2}{|c|}{\textbf{Gamma-ray quiet NLSy1s}} \\
\hline \ FBQS J1629+4007 \ & FBQS J1644+2619 \\
\hline \ SDSS J1722+5654 \ & SDSS J1450+5919 \\
\hline
\end{tabular}
\label{l2ea4-t1}
\end{center}
\end{table}

\section{Results}

\begin{center}
\begin{figure*}[!h]
\begin{center}
\begin{minipage}[t]{0.49\textwidth}
\begin{center}
\includegraphics[width=7cm,height=7cm]{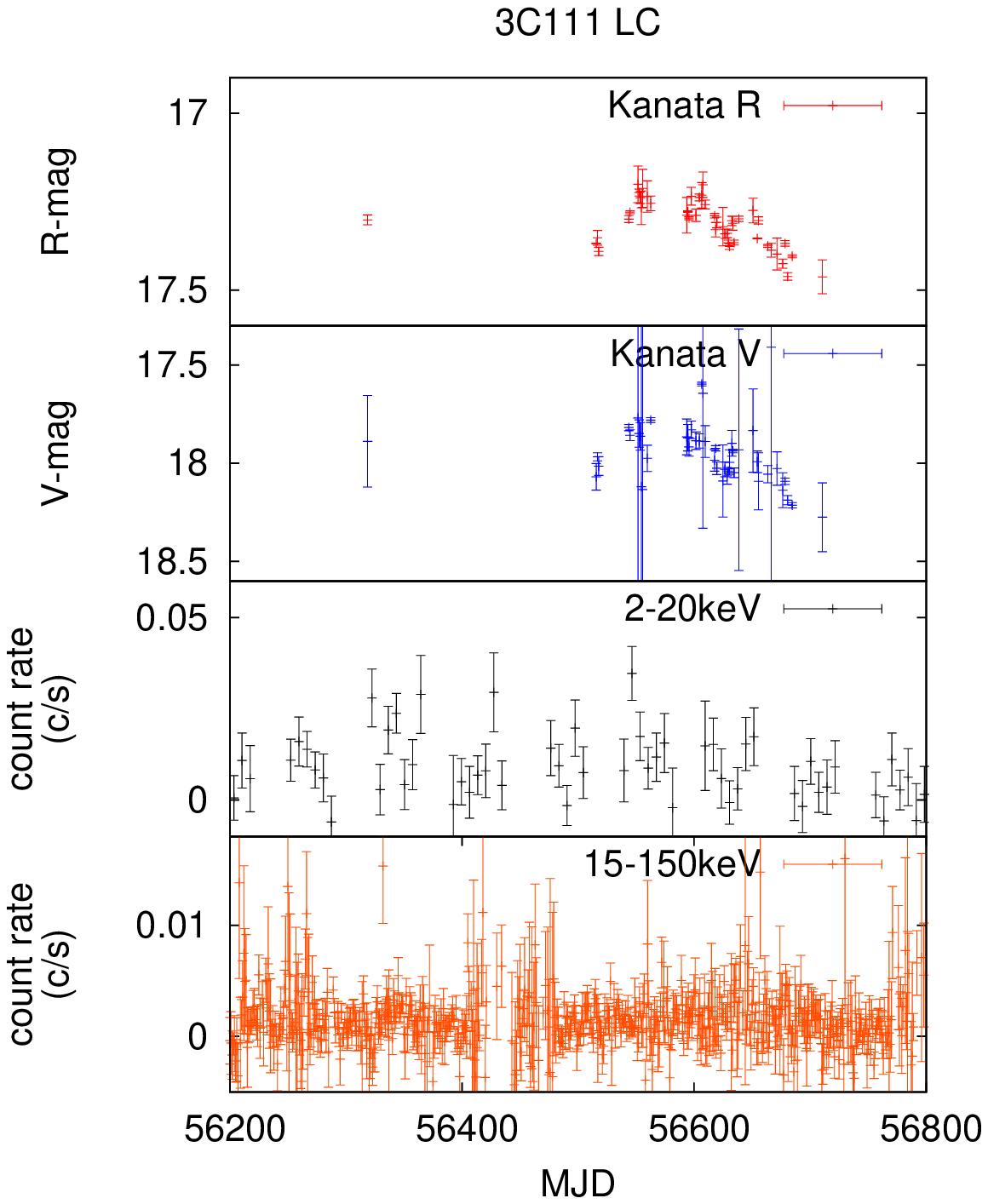}
\caption{Light curve of a radio galaxy 3C 111}
\label{fig:3c111}
\end{center}
\end{minipage}
\hfill
\begin{minipage}[t]{0.49\textwidth}
\begin{center}
\includegraphics[width=7cm,height=7cm]{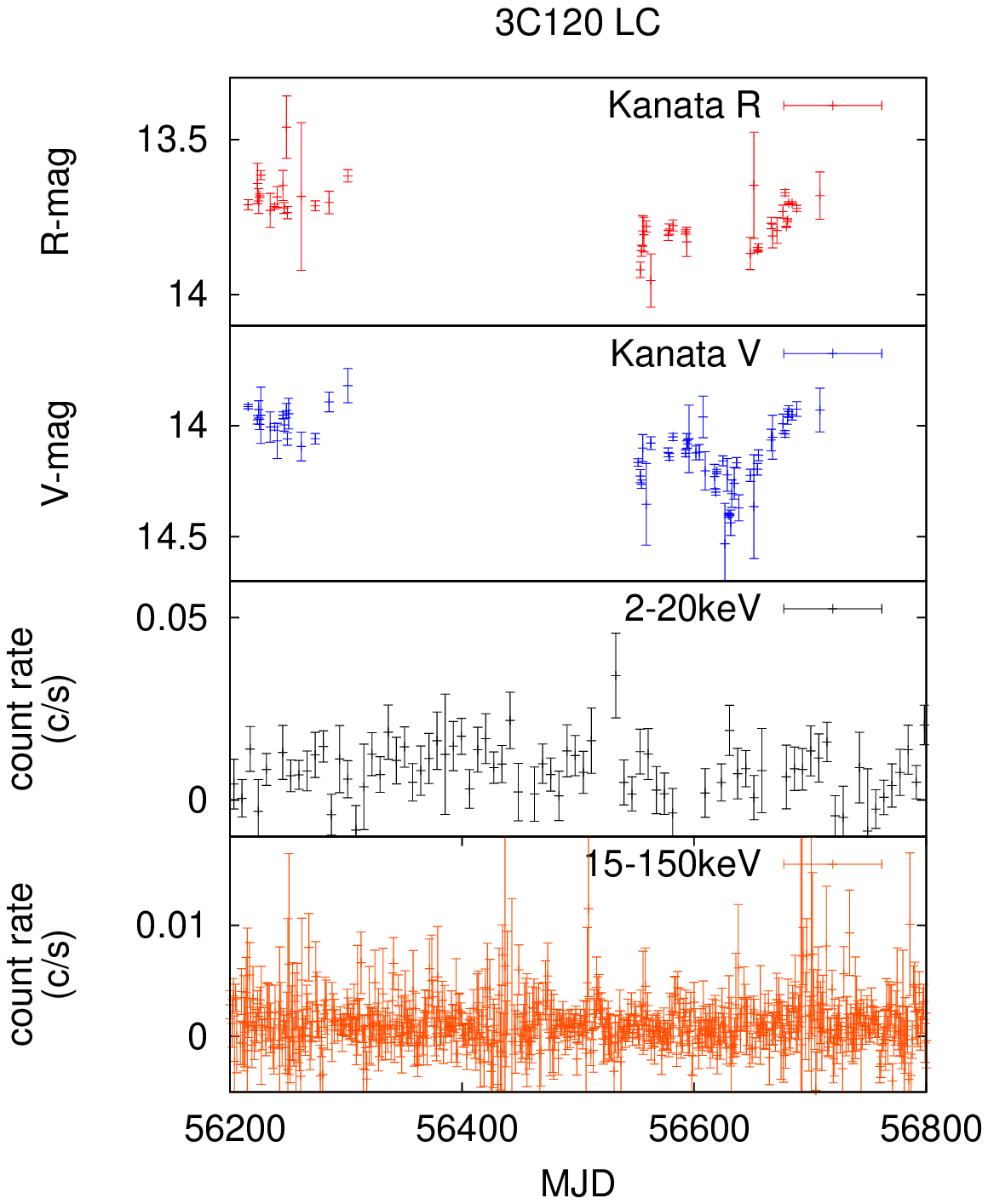}
\caption{Light curve of a radio galaxy 3C 120}
\label{fig:3c120}
\end{center}
\end{minipage}
\end{center}
%\end{figure*}
%\end{center}

%\begin{center}
%\begin{figure*}[!h]
\begin{center}
\begin{minipage}[t]{0.49\textwidth}
\begin{center}
\includegraphics[width=7cm,height=7cm]{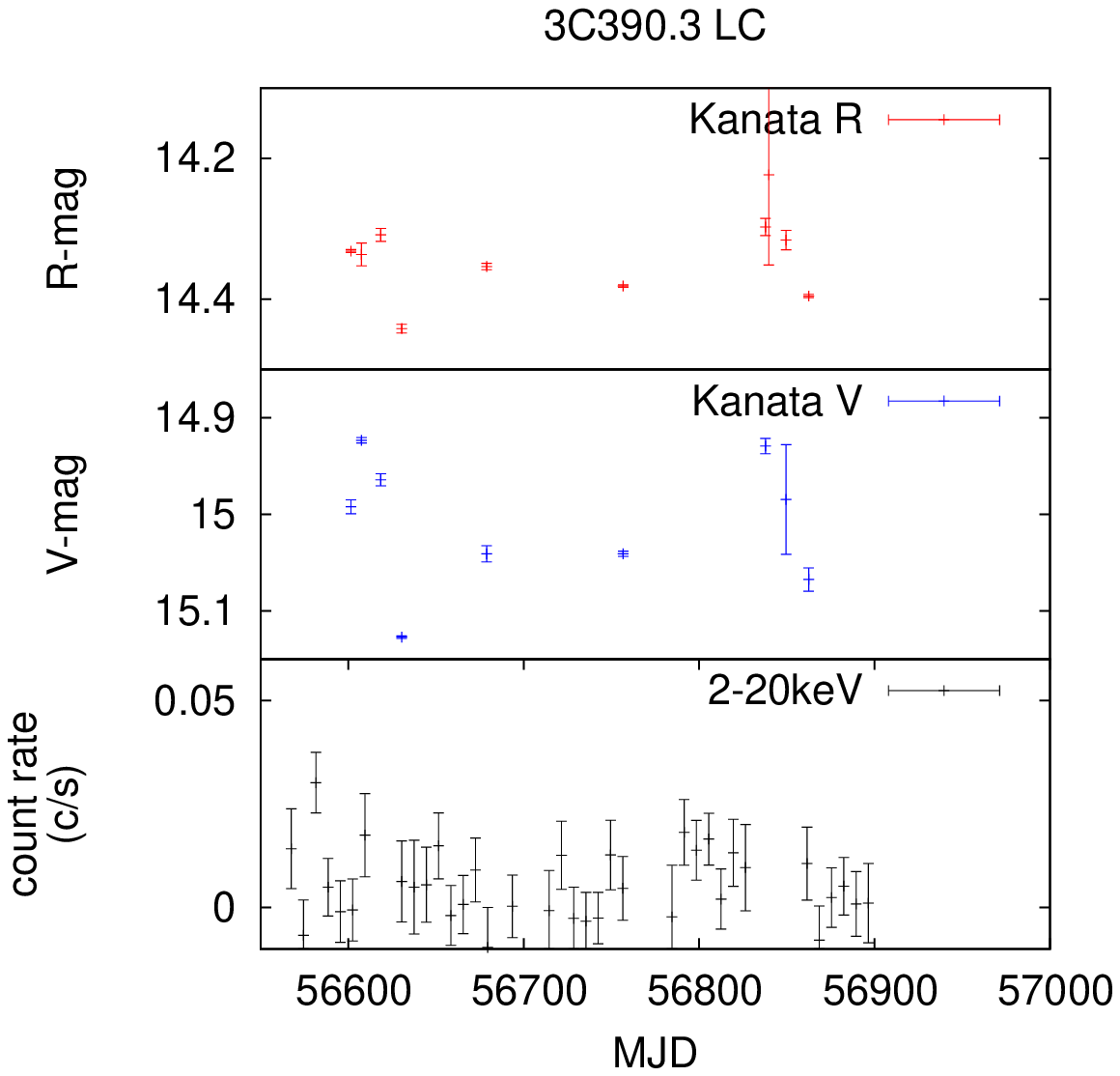}
\caption{Light curve of a radio galaxy 3C 390.3}
\label{fig:3c390}
\end{center}
\end{minipage}
\hfill
\begin{minipage}[t]{0.49\textwidth}
\begin{center}
\includegraphics[width=7cm,height=7cm]{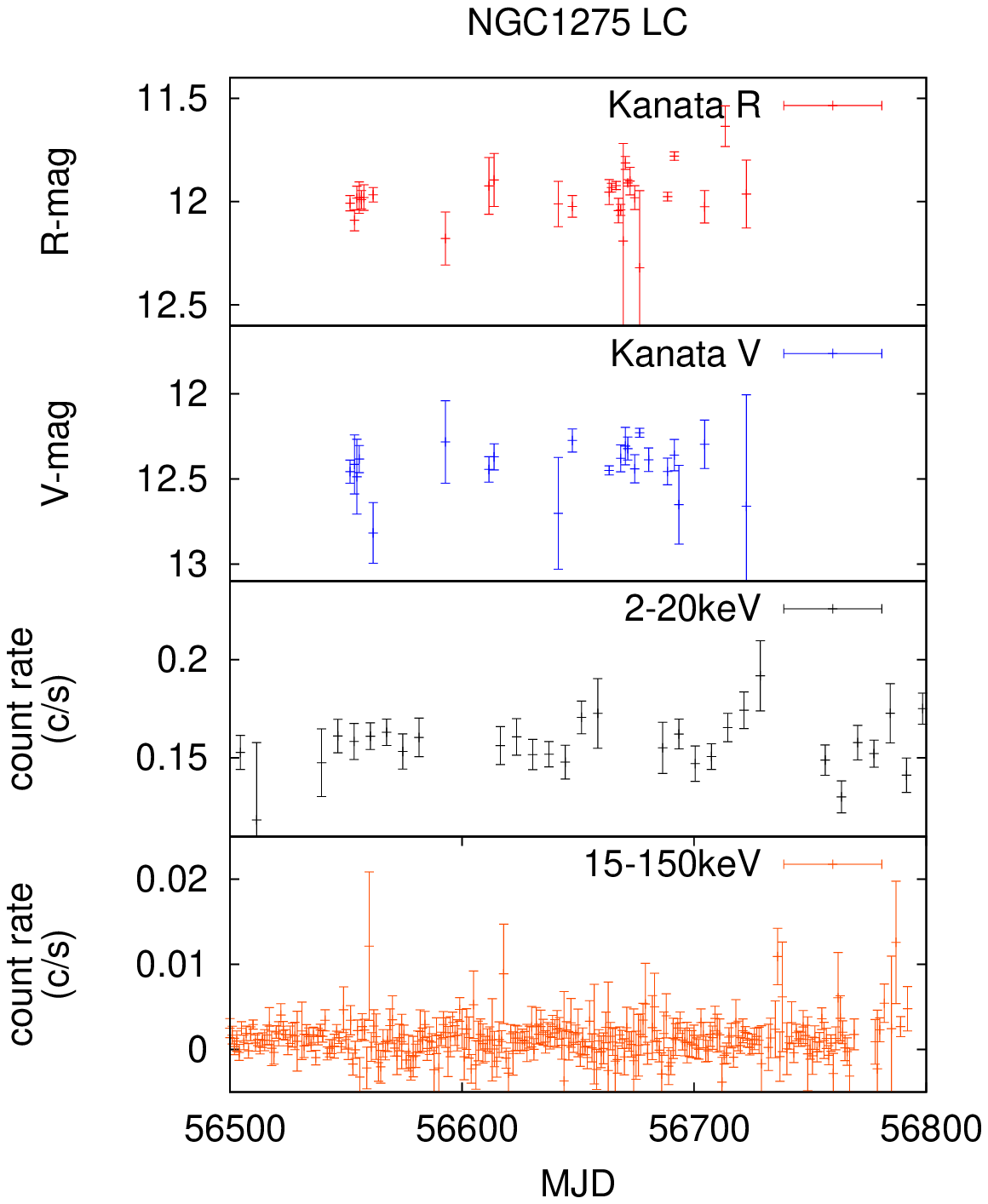}
\caption{Light curve of a radio galaxy NGC 1275}
\label{fig:ngc1275}
\end{center}
\end{minipage}
\end{center}
\end{figure*}
\end{center}

\textbf{Radio galaxies}\\
Fig \ref{fig:3c111}--\ref{fig:ngc1275} show the results for radio galaxies. Each figure show
optical R-band(top), V-band(second) magnitude by Kanata, 
2-20 keV daily X-ray count rate by
MAXI (third), and 15-150 keV weekly count rate by Swift-BAT (bottom). 
The gaps in MAXI light curves are the period when objects are
not in FOV of MAXI. There is no Swift-BAT public data for 3C 390.3.  
We can see a clear flux variability in the optical band for
3C 111 and 3C 120, but no object shows a significant X-ray flux variability.

\begin{center}
\begin{figure*}[!h]
\begin{center}
\begin{minipage}[t]{0.49\textwidth}
\begin{center}
\includegraphics[width=7cm,height=7cm]{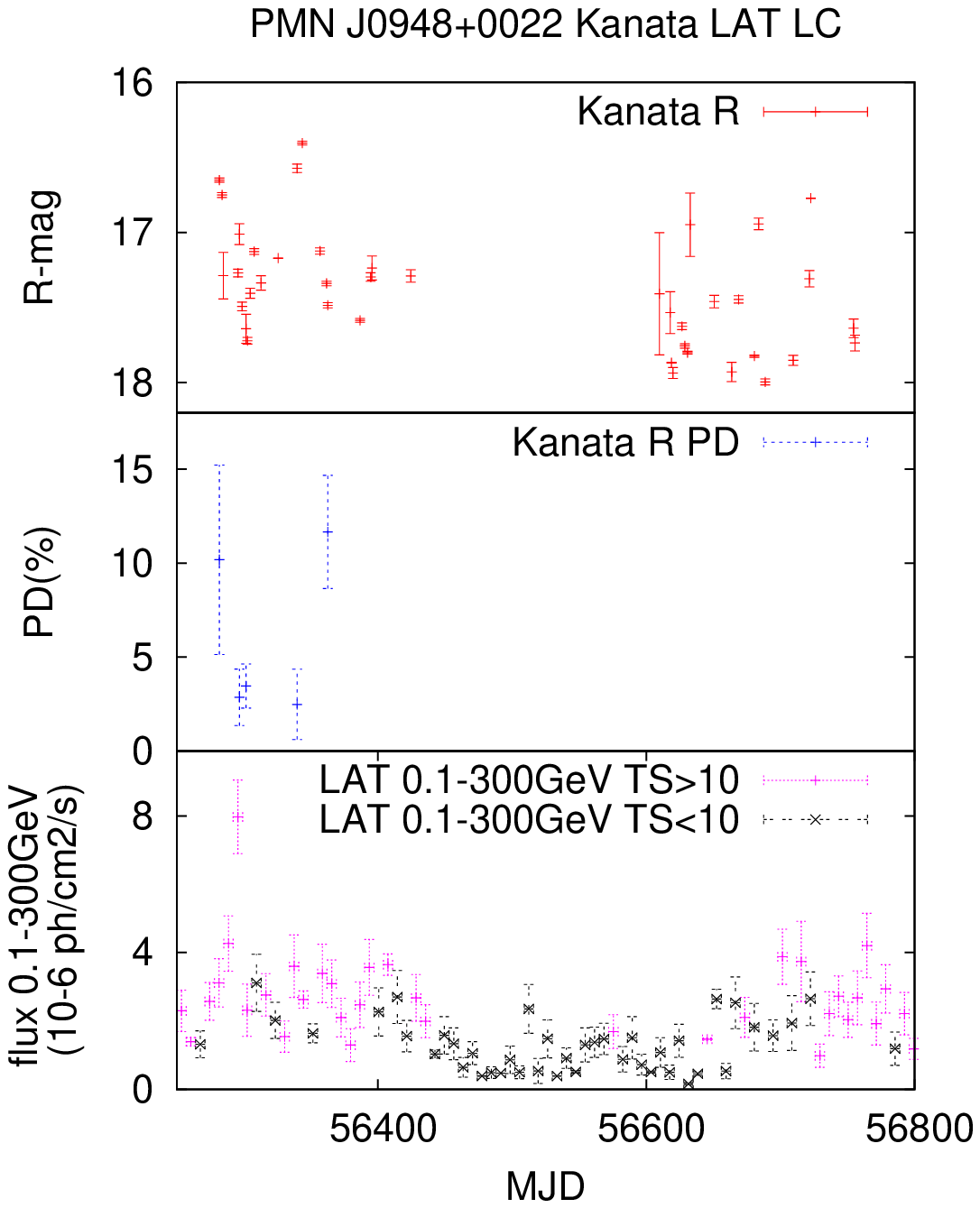}
\caption{Light curve of a gamma-ray loud NLSy1 PMN J0948+0022}
\label{fig:qso0948_all}
\end{center}
\end{minipage}
\hfill
\begin{minipage}[t]{0.49\textwidth}
\begin{center}
\includegraphics[width=7cm,height=7cm]{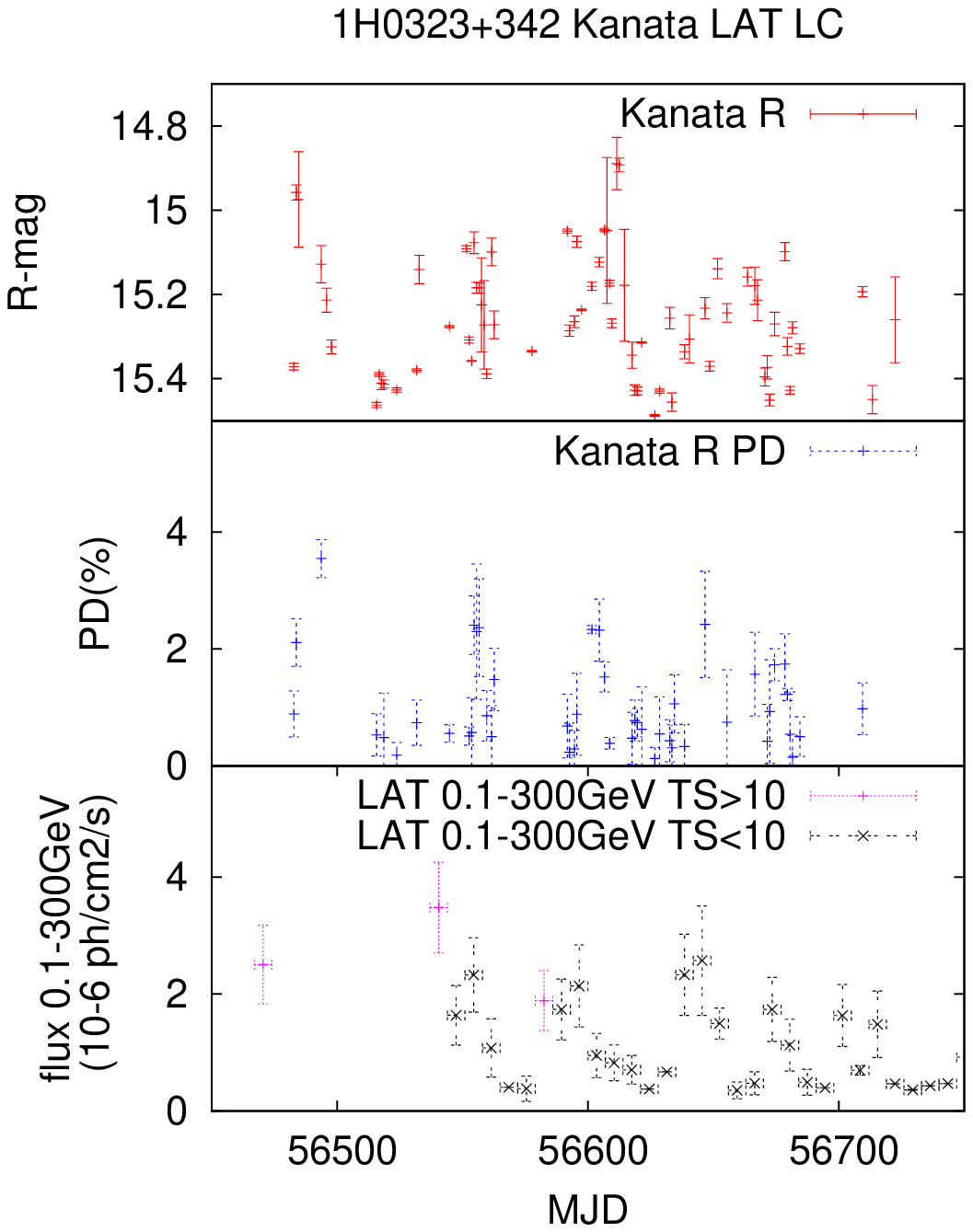}
\caption{Light curve of a gamma-ray loud NLSy1 1H 0323+342}
\label{fig:1h0323_all}
\end{center}
\end{minipage}
\end{center}
%\end{figure*}
%\end{center}

%\begin{center}
%\begin{figure*}[!h]
\begin{center}
\begin{minipage}[t]{0.49\textwidth}
\begin{center}
\includegraphics[width=7cm,height=7cm]{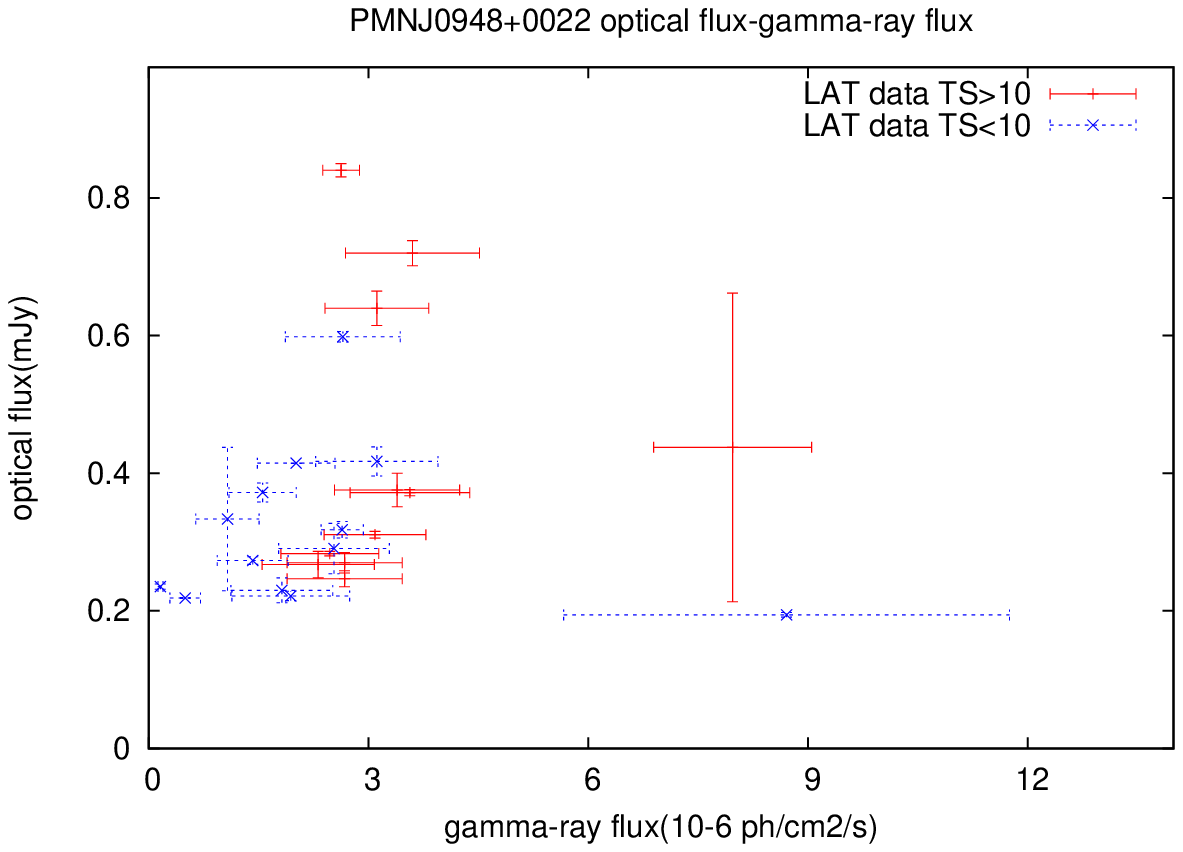}
\caption{Correlation between optical flux and gamma-ray flux of PMN J0948+0022}
\label{fig:qso0948_optflux_gammaflux}
\end{center}
\end{minipage}
\hfill
\begin{minipage}[t]{0.49\textwidth}
\begin{center}
\includegraphics[width=7cm,height=7cm]{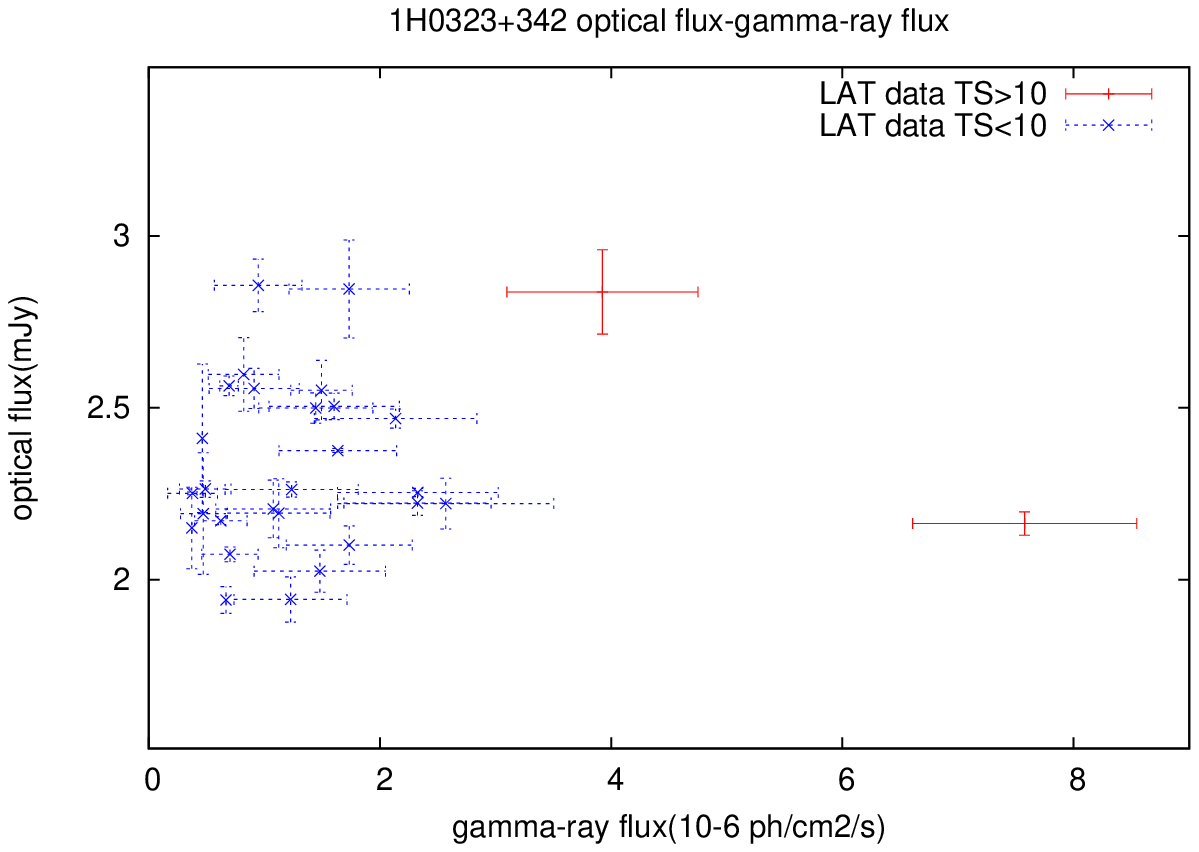}
\caption{Correlation between optical flux and gamma-ray flux of 1H 0323+342}
\label{fig:1h0323_optflux_gammaflux}
\end{center}
\end{minipage}
\end{center}
%\end{figure*}
%\end{center}

%\begin{center}
%\begin{figure*}[!h]
\begin{center}
\begin{minipage}[t]{0.49\textwidth}
\begin{center}
\includegraphics[width=7cm,height=7cm]{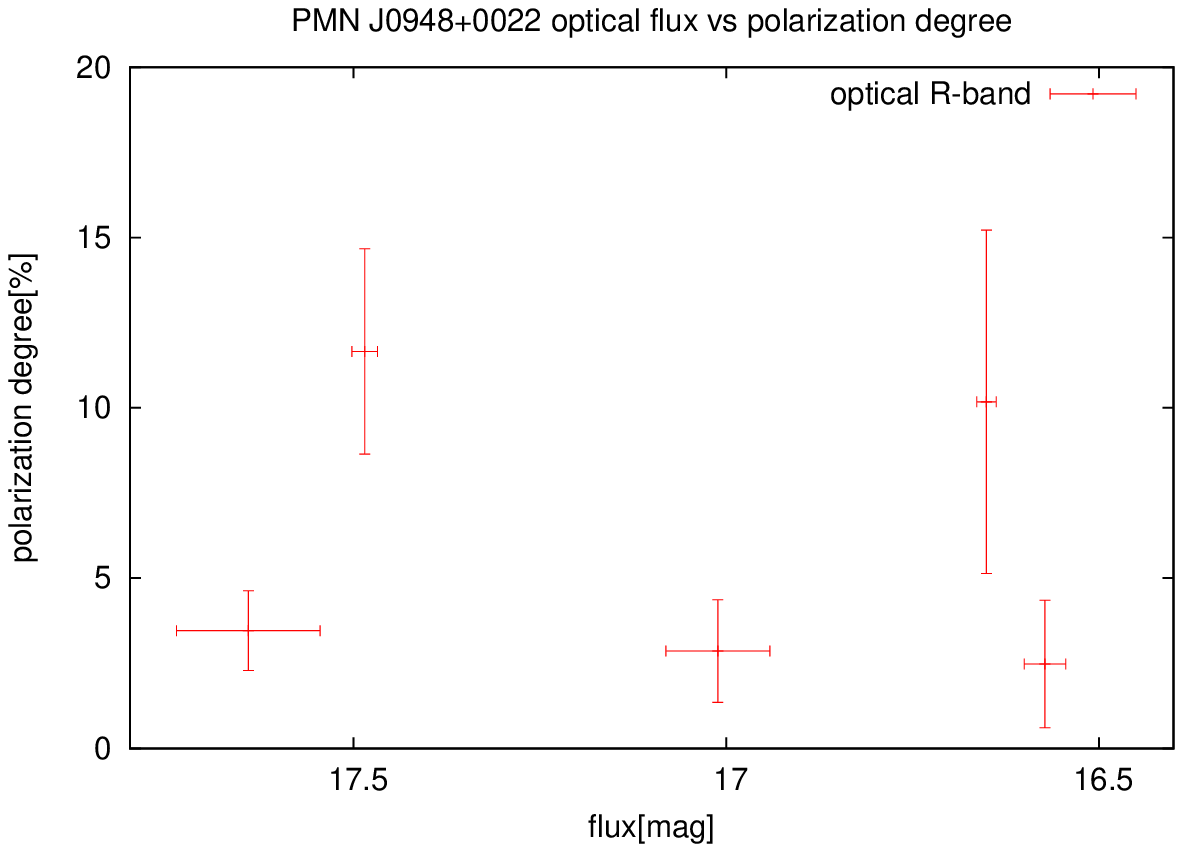}
\caption{Correlation between optical flux and optical polarization
 degree of PMN J0948+0022}
\label{fig:qso0948_optflux_optpol}
\end{center}
\end{minipage}
\hfill
\begin{minipage}[t]{0.49\textwidth}
\begin{center}
\includegraphics[width=7cm,height=7cm]{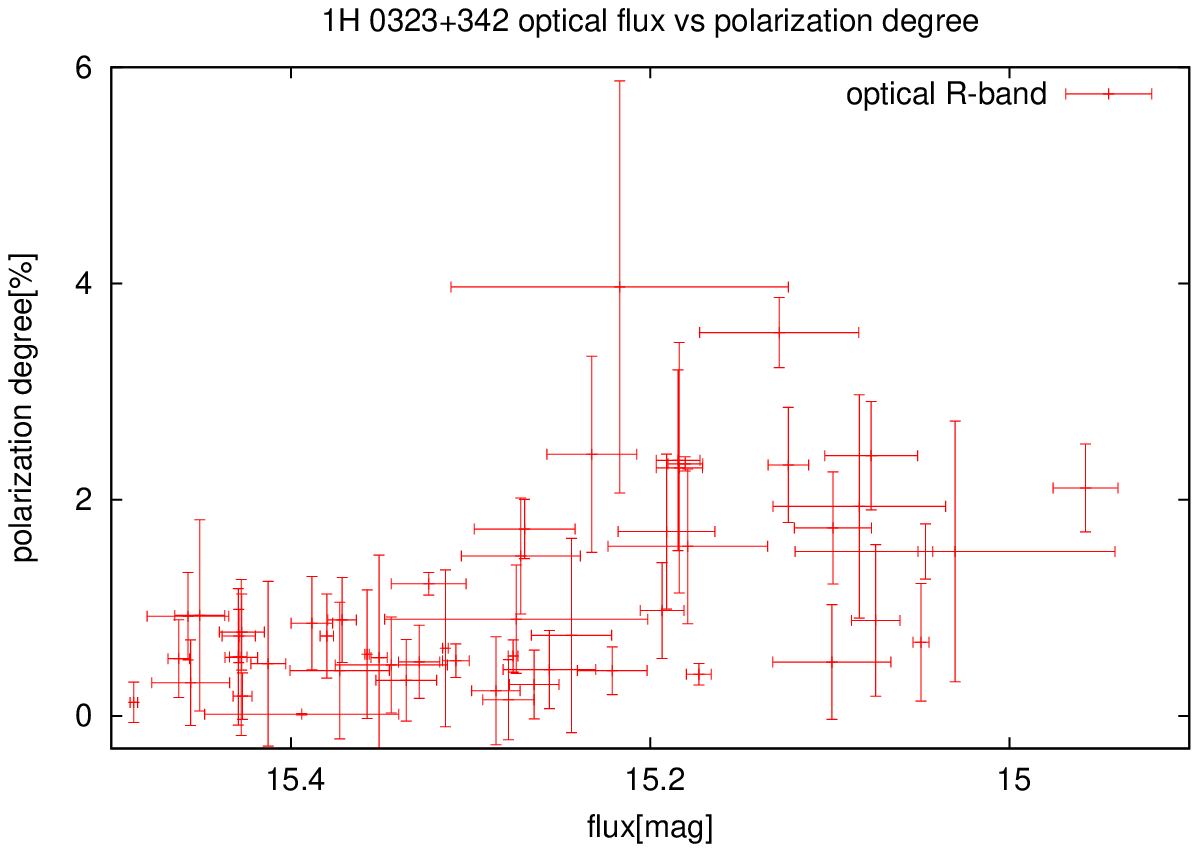}
\caption{Correlation between optical flux and optical polarization
 degree of 1H 0323+342}
\label{fig:1h0323_optflux_optpol}
\end{center}
\end{minipage}
\end{center}
\end{figure*}
\end{center}

\textbf{Gamma-ray loud NLSy1s}\\
Fig \ref{fig:qso0948_all} and \ref{fig:1h0323_all} show the light curves of gamma-ray loud NLSy1. Each
figure shows optical R-band (top), optical R-band polarization degree
(middle), and 0.1-300 GeV
gamma-ray (bottom). We can see some flares in optical and gamma-ray bands.
Fig \ref{fig:qso0948_optflux_gammaflux} and \ref{fig:1h0323_optflux_gammaflux} show the correlation between optical
flux and gamma-ray flux. We cannot see any clear correlation between these
bands.\\
Fig \ref{fig:qso0948_optflux_optpol} and \ref{fig:1h0323_optflux_optpol}
show the correlation between flux and polarization
degree (PD) in the optical band. In PMN J0948+0022, optical flux and PD are not
correlated and maximum PD reaches more than 10\%. 1H 0323+342 shows high optical PD when optical flux is high. But
optical PD value is low in the whole period (less than 5\%).

\begin{center}
\begin{figure*}[!h]
\begin{center}
\begin{minipage}[t]{0.49\textwidth}
\begin{center}
\includegraphics[width=7cm,height=7cm]{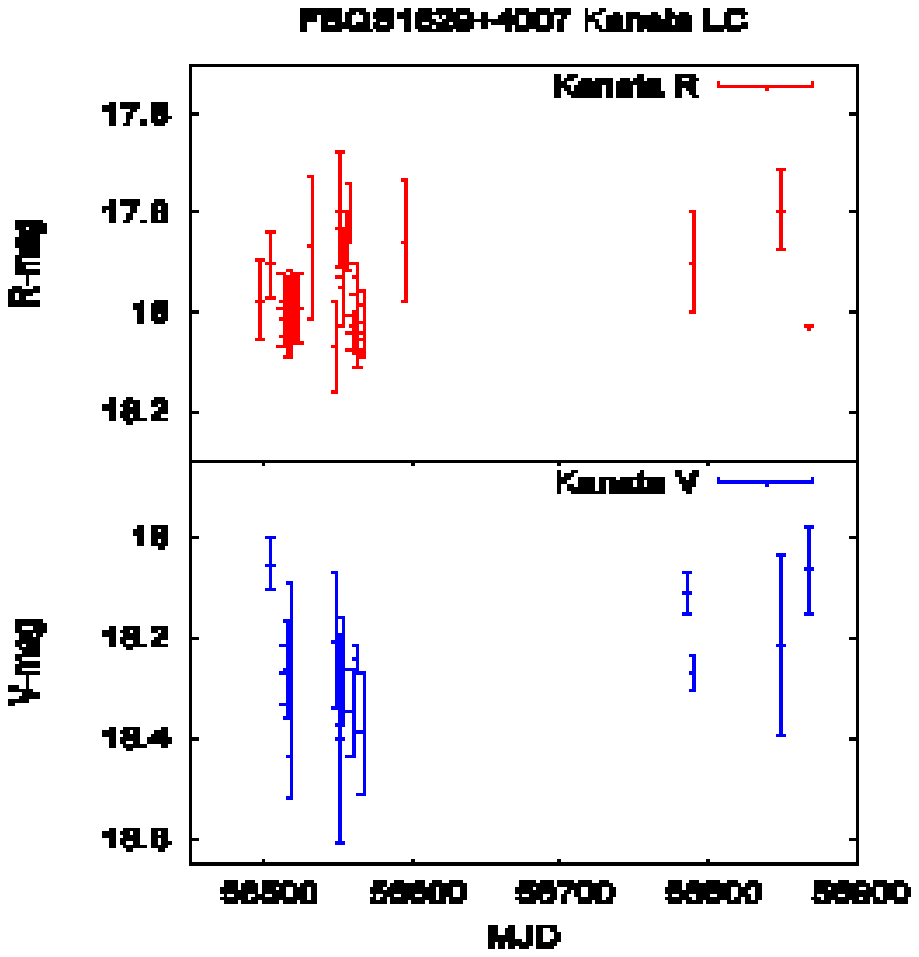}
\caption{Light curve of a gamma-ray quiet NLSy1 FBQS J1629+4007}
\label{fig:1629_lc}
\end{center}
\end{minipage}
\hfill
\begin{minipage}[t]{0.49\textwidth}
\begin{center}
\includegraphics[width=7cm,height=7cm]{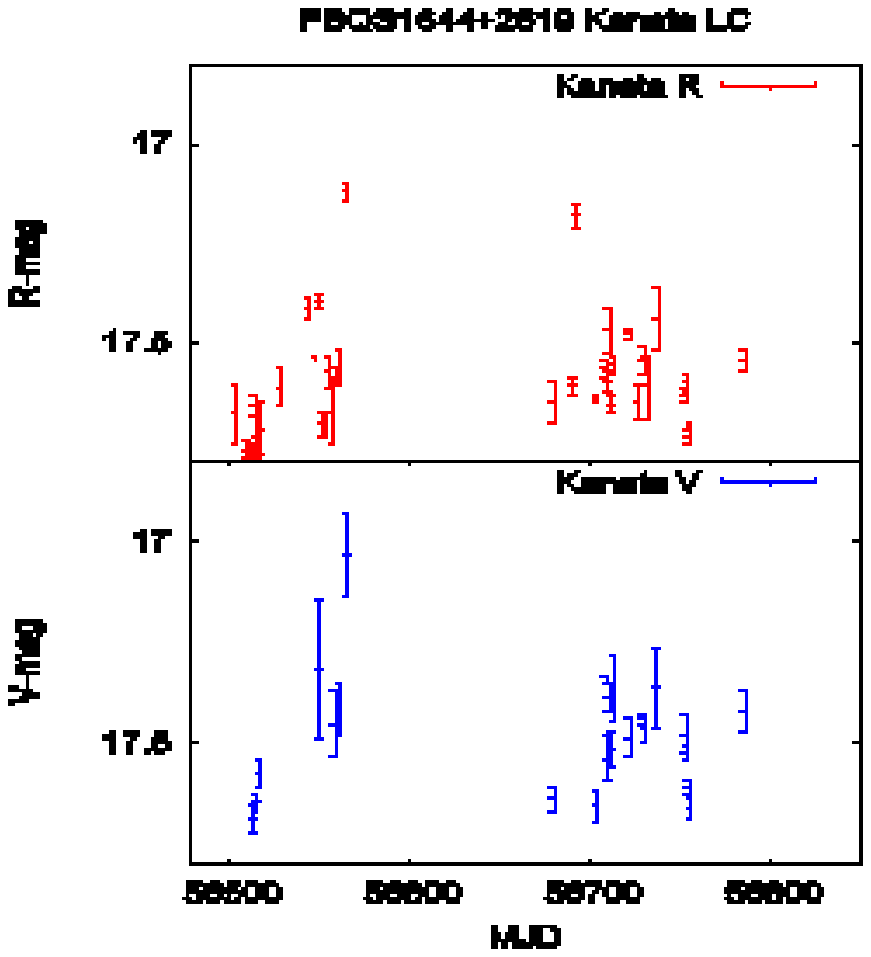}
\caption{Light curve of a gamma-ray quiet NLSy1 FBQS J1644+2619}
\label{fig:1644_lc}
\end{center}
\end{minipage}
\end{center}
%\end{figure*}
%\end{center}

%\begin{center}
%}\begin{figure*}[!h]
\begin{center}
\begin{minipage}[t]{0.49\textwidth}
\begin{center}
\includegraphics[width=7cm,height=7cm]{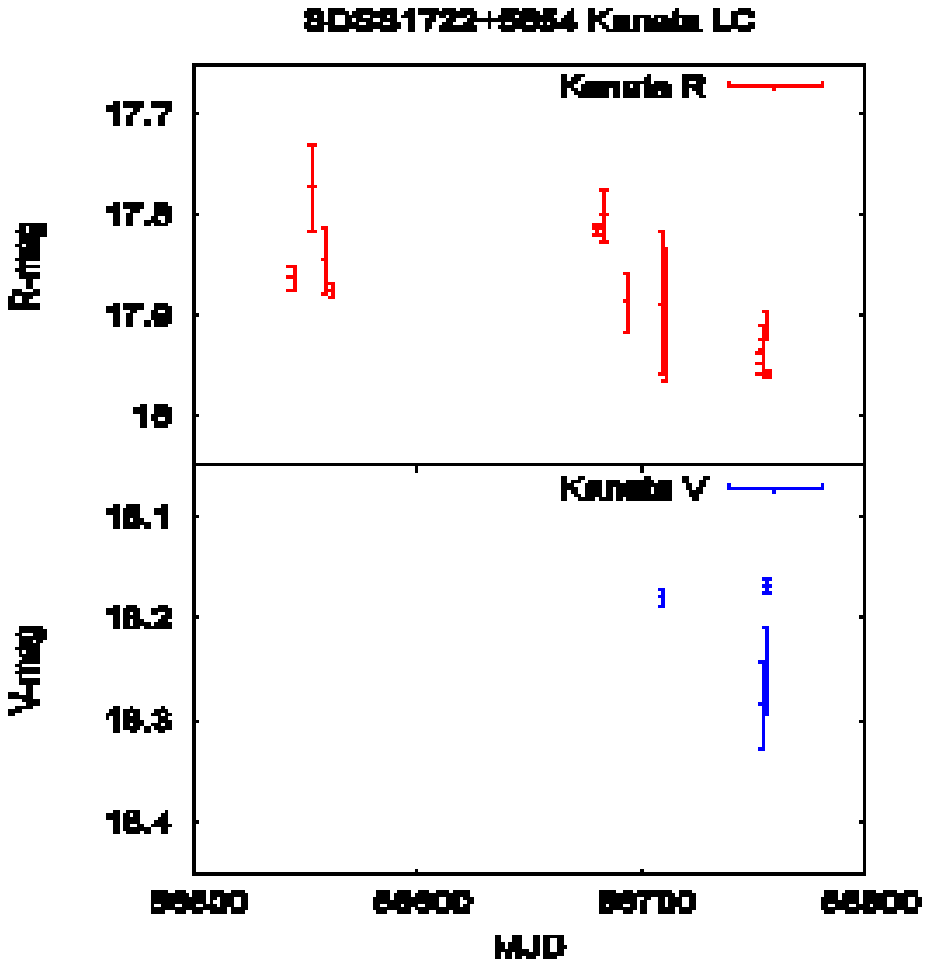}
\caption{Light curve of a gamma-ray quiet NLSy1 SDSS J1722+5654}
\label{fig:1722_lc}
\end{center}
\end{minipage}
\hfill
\begin{minipage}[t]{0.49\textwidth}
\begin{center}
\includegraphics[width=7cm,height=7cm]{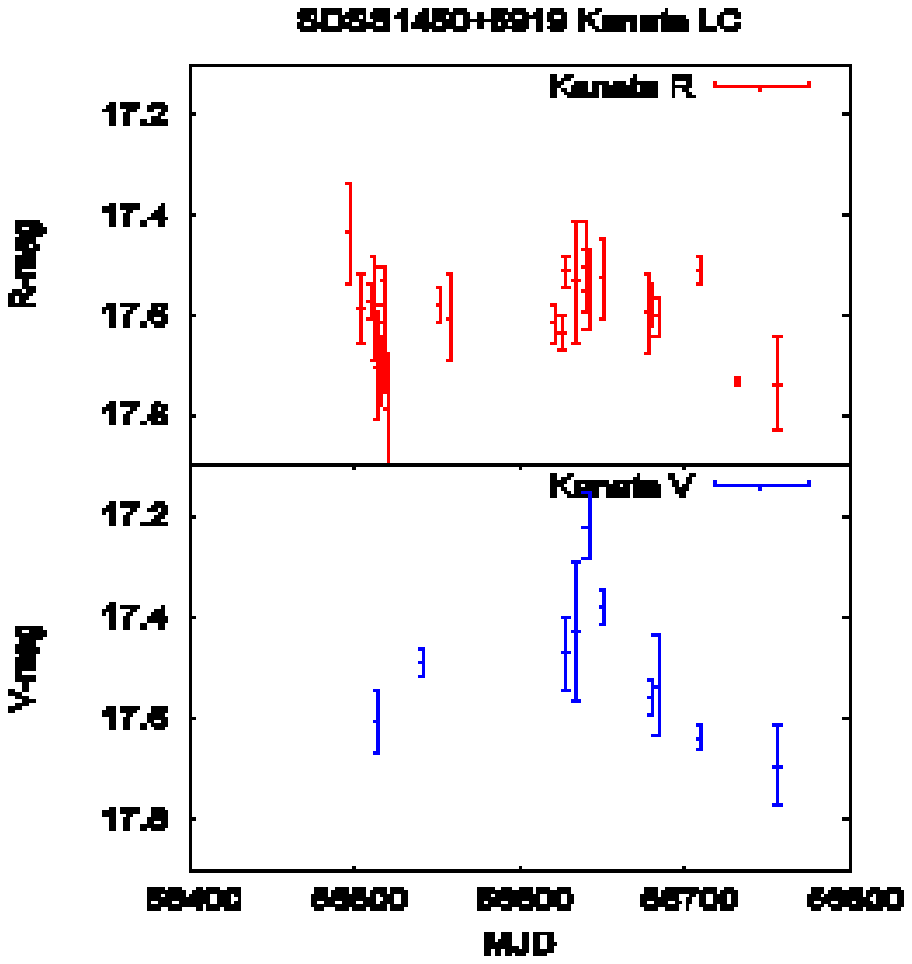}
\caption{Light curve of a gamma-ray quiet NLSy1 SDSS J1450+5919}
\label{fig:1450_lc}
\end{center}
\end{minipage}
\end{center}
\end{figure*}
\end{center}

\textbf{Gamma-ray quiet NLSy1s}\\
Fig \ref{fig:1629_lc}--\ref{fig:1450_lc} shows the results for gamma-ray quiet NLSy1. Each
figure show optical R-band light curve (top) and V-band light curve
(bottom). Only FBQS J1644+2619 shows daily-scale flux variability around
MJD=56560. In this period, optical flux increases about 0.4 mag in 3 days both in R-band and V-band.
\\

\section{Discussion}
\begin{center}
\begin{figure}[!h]
\includegraphics[width=10cm]{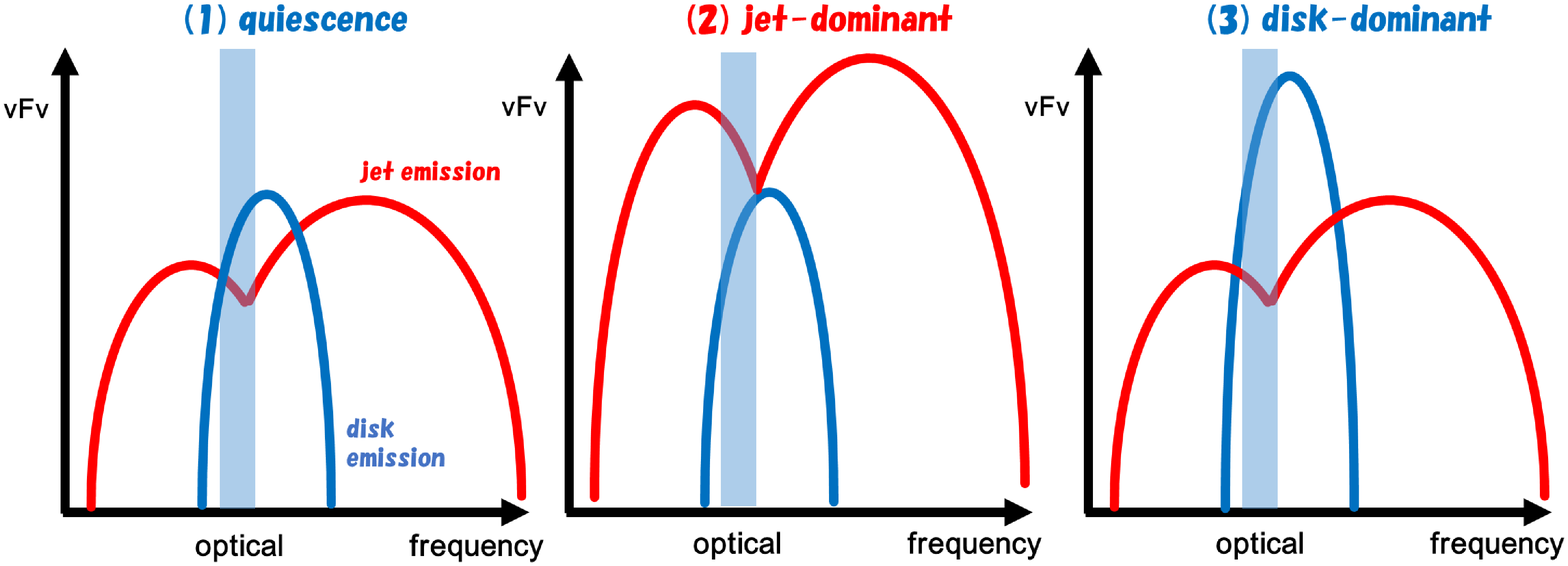}
\caption{optical emission phases for RL-NLSy1s}
\label{fig:emission_phase}
\end{figure}
\end{center}

No radio galaxies show a clear correlation between optical and X-ray
bands.
Probably this is due to low signal-to-noise ratio of X-ray
light curves.
We need more sensitive monitoring for X-ray study.
\\
For gamma-ray loud NLSy1s, a jet-dominant phase or a disk-dominant
phase is inferred to appear in addition to quiet phase. As shown in Fig
\ref{fig:emission_phase}, (1) in quiescence, disk emission seems to be
dominant in optical band. 
(2) Jet-dominant phase is suggested to appear as flares in both
optical and gamma-ray with an increase of an optical polarization degree. 
(3) Disk-dominant phase is indicated to be a quiet phase or an optical flare
without polarization increase. 
Optical polarization and variability time-scale
are important information to study emission mechanism in the optical
band. Jet emission is polarized and shows a short-term variability.
Further dense monitoring observations are needed to conclude the above.
\\
For gamma-ray quiet NLSy1s, 
only FBQS J1644+2619 shows rapid flux variability.
Previous studies in the radio band for FBQS
J1644+2619 shows characteristics similar to
blazars. Also, this object is listed in the 3rd FGL catalog as a new
gamma-ray source. Optical short-term variability supports that this
object shows a synchrotron emission in the optical band during the flare,
like Fig. \ref{fig:emission_phase} (2) jet-dominant phase.
\\

%%\bigskip % extra skip inserted
%\begin{acknowledgments}

%\end{acknowledgments}

%%\bigskip % extra skip inserted

\end{document}